\documentclass[epj]{svjour}
\usepackage{graphics}
\usepackage{graphicx}
\usepackage{textcomp}
\usepackage{makeidx}
\usepackage{amsmath}
\usepackage{subfigure}
\usepackage{amssymb}
\usepackage{hyperref}
\usepackage{graphicx}
\usepackage{color}
\usepackage{float}
\usepackage{cite}
\usepackage{epstopdf}
\begin{document}
\title{A way of decoupling gravitational sources in Pure Lovelock gravity}
\titlerunning{A way of decoupling gravitational sources in pure Lovelock gravity }
\author{Milko Estrada\inst{1}  \thanks{\emph{e-mail:} milko.estrada@gmail.com} }
\authorrunning{M Estrada}
%

%
\institute{Departamento de F\'isica, Facultad de ciencias b\'asicas, Universidad de Antofagasta, Casilla 170, Antofagasta, Chile. }

\date{Received: date / Revised version: date}
%
\abstract{ We provide an algorithm that shows how to decouple gravitational sources in Pure Lovelock gravity. 
This method allows to obtain several new and known analytic solutions of physical interest in scenarios with extra dimensions and with presence of higher curvature terms. Furthermore, using our method, it is shown that applying the minimal geometric deformation to the Anti de Sitter space time it is possible to obtain regular black hole solutions.}  

\PACS{
      {PACS-key}{discribing text of that key}   \and
      {PACS-key}{discribing text of that key}
     } 
%

\maketitle
\section{Introduction}
In the last years, several branches of theoretical physics have predicted the presence of extra dimensions. Thus, it makes sense to think of the existence of geometrical objects in space times with a number of dimensions greater than four as for example branes, strings, or higher dimensional black holes. 
In regarding this, theories of gravity emerged which present higher curvature correction terms when the space time has a number of dimensions greater than four. One interesting theory among of them is {\it Lovelock gravity} \cite{Lovelock}. One achievement of Lovelock theory is that it shares the following features with the General Relativity :

\begin{itemize}
    \item Its equation of motion are symmetric. 
    \item Its equations of motion are of second order on the derivative of the metric tensor.
    \item Free divergence.
\end{itemize}

Into the Lovelock gravities, we can find the {\it Pure Lovelock theory}. It is well known that the General Relativity has a no non-trivial vacuum solution (without cosmological constant) when $d=3$ ({\it i.e} $d=2n+1$, where $n=1$), one interesting feature is that Pure Lovelock keeps this property for $d=2n+1$ with $n>1$, see reference \cite{indio6}. On the other hand, including the cosmological constant, General Relativity has a unique (Anti) de Sitter {\it ground state} for $\Lambda (<0)>0$, in regarding this, other interesting feature of Pure Lovelock theory is that, it keeps this feature for $n$ odd, however, for $n$ even, this theory has a double Anti de Sitter or de Sitter ground state for $\Lambda>0$ and does not have ground state for $\Lambda<0$ \cite{milko1,Cai:2006pq}.

On the other hand, in Einstein Hilbert theory, finding new solutions of physical interest is not a simple task due to the highly nonlinear behavior of its equations. In regarding this, Ovalle in (2017) \cite{Ovalle1} proposed a method called {\it Gravitational Decoupling of Sources}, which corresponds to the first algorithm that shows how to decouple gravitational sources in General Relativity. This method applies a {\it Minimal Geometric Deformation} (MGD) to the temporal and radial metric components together with a {\it decoupling of sources}. The method is explained  in reference \cite{Ovalle2}: ``given two gravitational sources: a source A and an extra source B, standard Einstein’s equations are first solved for A, and then a simpler set of {\it quasi-Einstein} equations are solved for B. Finally, the two solutions can be combined in order to derive the complete solution for the total system.".

By applying this method to a solution of Einstein equations, named {\it seed solution}, it is possible to obtain new analytic solutions of physical interest. Related to this, by deforming some isotropic well known and well behaved solutions, new anisotropic and well behaved solutions have been obtained that represent stellar distributions in references \cite{Ovalle2,Camilo, Tello1, Graterol,Tello2,Tello3,Luciano,Milko}. By deforming the Schwarzschild space time new black hole solutions have been obtained in reference \cite{Ovalle3} ( other black hole solutions obtained with this method in references \cite{Contreras1,Contreras2,Contreras3,Contreras4} ). Other examples of applications of the method are: solutions in Einstein Klein Gordon system \cite{Ovalle4}; solutions in $f(G)$ gravity \cite{Sharif1}; solutions in $f(R)$ gravity \cite{Sharif5},cloud of strings solutions \cite{Angel}. See other applications in references \cite{Ovalle5,Contreras6,Sharif2,Sharif3,Luciano1,Sharif4,Regginaldo,Ovalle:2019lbs,Ovalle:2019Iso,Gabbanelli:2019txr}.  

Thus, motivated by the fact that in Pure Lovelock theory there are several kind of solutions in literature as for example: black hole solutions in references \cite{Cai:2006pq,indio3,indioBH,PureLovelock1,PureLovelock2, Aranguiz:2015voa} and stellar distributions in references \cite{indio2,indio4,indio5}, it seems of physical interest to provide an algorithm to decouple gravitational sources in Pure Lovelock gravity, and thus, to apply the method to known ( or unknown) solutions and to test if it is possible to obtain new solutions of physical interest. See also \cite{milko1,milko2,indio1}.

We start by deforming the seed energy momentum tensor $\bar{T}_{A B}$ by an additional source $\theta_{A B}$, which causes anisotropic effects on the self-gravitating system. This additional source can contain new fields, like scalar, vector and tensor fields \cite{Ovalle2}. Therefore the energy momentum tensor is:

\begin{equation}
    T_{A B} = \bar{T}_{A B} + \theta_{A B}, \label{EM}
\end{equation}
since Pure Lovelock theory has free divergence, then, the energy momentum tensor satisfies the conservation equation:
\begin{equation}
    \nabla_A T^{A B} =0.
\end{equation}

In this work we provide a Gravitational Decoupling method in Pure Lovelock gravity, and therefore we show a simple approach to decoupling gravitational sources in this theory. We will show that the Pure Lovelock equations of motion can be solved for each component $\{\bar{T}_{A B},\theta_{A B}\}$ separately, at least for the spherically symmetric and static case. For each component will be obtained a particular metric tensor $\{\bar{g}_{A B}, g_{ A B} ^{\,\,\,\theta} \}$, and the final metric $g_{A B}$ is a simple combination of these metrics. As a simple test, we will apply our method to an Anti de Sitter space time and, we will test if it is possible to obtain solutions that represent regular black holes.

\section{Lovelock Gravity and the Pure Lovelock case} \label{Lovelock}
The Lovelock Lagrangian is :
\begin{equation}\label{LovelockLagrangian}
L  = \sqrt{-g} \sum_{n=0}^N \gamma_n L_n,
\end{equation}
where $N=\frac{d}{2}-1$  for $d$ even and $N=\frac{d-1}{2}$ for $d$ odd and, $\gamma_n$ are arbitrary coupling constants. $L_n$ is a topological density defined as:
\begin{equation}
 L_n = \frac{1}{2^n} \delta^{\mu_1 \nu_1 ...\mu_n \nu_n}_{\alpha_1 \beta_1 ... \alpha_n \beta_n} \displaystyle \Pi^n_{r=1} R^{\alpha_r \beta_r}_{\mu_r \nu_r},
\end{equation}
where $R^{\alpha \beta}_{\mu \nu}$ is a $n$ order generalization of the Riemann tensor for the Lovelock theory, and:
\begin{equation}
 \delta^{\mu_1 \nu_1 ...\mu_n \nu_n}_{\alpha_1 \beta_1 ... \alpha_n \beta_n} = \frac{1}{n!} \delta^{\mu_1}_{\left[\alpha_1\right.} \delta^{\nu_1}_{\beta_1} ... \delta^{\mu_n}_{\alpha_n} \delta^{\nu_n}_{\left.\beta_n \right]}   
\end{equation}
is the {\it generalized Kronecker delta}.

It is worth to stress that, the terms $L_0,L_1$ and $L_2$ are proportional to the cosmological constant, Ricci Scalar and the Gauss Bonnet Lagrangian, respectively. The corresponding equation of motion is given by: 

\begin{equation}
 \sum ^N_{n=0} \gamma_n \mathcal{G}^{(n)}_{AB} =   T_{AB},
\end{equation}
where $\mathcal{G}^{(n)}_{AB}$ is a $n$ order generalization of the Einstein tensor due to the topological density $L_n$. As example $\mathcal{G}^{(1)}_{AB}$ is just the Einstein tensor associated with the Ricci scalar (Einstein Hilbert theory is a particular case of Lovelock theory), and $\mathcal{G}^{(2)}_{AB}$ is the Lanczos tensor $H_{AB}$ associated with the Gauss Bonnet Lagrangian. 

For example, the Einstein Gauss Bonnet equations of motion up to $n=2$, without cosmological constant are:

\begin{equation}
G^{A}_{B} + \gamma_2 H^{A}_{B} = T^{A}_{B}  
\end{equation}
where the Lanczos tensor is:
\begin{align}
H_{AB} =& 2\Big (R R_{AB} - 2R_{AC}R^C_B - 2R^{CD}R_{ACBD}  \nonumber \\
        &+ R^{CDE}_{A} R_{BCDE} \Big)- \frac{1}{2} g_{AB} L_2.
\end{align}

\subsection{Pure Lovelock case} \label{PureLovelock}
Pure Lovelock is a theory that involving only a single fixed value of $n$ (with $n \ge 1$),  without sum over the lower order. In some cases it is considered one single value of $n \ge 1$ plus the $n=0$ term, {\it i.e.} $L=L_0+L_n$, as for example in references \cite{milko1,PureLovelock1,PureLovelock2} . For simplicity, in this work we take a single value of $L_n$ without the $L_0$ term, as for example in references \cite{indio1,indio2} . Thus, the Lagrangian is:
\begin{equation}
 L = \sqrt{-g} \gamma_n  L_n= \sqrt{-g} \gamma_n \frac{1}{2^n} \delta^{\mu_1 \nu_1 ...\mu_n \nu_n}_{\alpha_1 \beta_1 ... \alpha_n \beta_n} \displaystyle \Pi^n_{r=1} R^{\alpha_r \beta_r}_{\mu_r \nu_r},
\end{equation}

The equations of motion are given by:

\begin{equation} \label{eqmovimiento}
 \mathcal{G}^{(n)}_{AB} = T_{AB} , 
\end{equation}
where
\begin{equation}
    (\mathcal{G}^{(n)})_{B}^A=-\frac{1}{2^{n+1}}\,\delta_{B A_{1}...A_{2n}}^{A
B_{1}...B_{2n}}\,R_{B_{1} B_{2}}^{A_{1} A_{2}}\cdots R_{B
_{2n-1} B_{2n}}^{A_{2n-1} A_{2n}}\,.
\end{equation}
 and where the coupling constants were set to unity as in references \cite{indio1,indio2}.
 
In this work we study the static $d$ dimensional spherically symmetric metric, wich in Schwarzschild-like coordinates reads:
\begin{equation}
    ds^2=-e^{\nu (r)}+e^{\lambda (r)} dr^2+r^2 d\Omega^2_{d-2}, \label{metrica1}
\end{equation}
where $d\Omega^2_{d-2}$ corresponds to the metric of a $(d-2)$ unitary sphere. The energy momentum tensor corresponds to a  neutral perfect fluid:

\begin{equation}
T^A_B=\mbox{diag}(-\rho,p_r,p_\theta,p_\theta,...),    
\end{equation}
where, from the spherical symmetry, we have for all the $(d-2)$ angular coordinates that $p_\theta=p_\phi=...$ . The conservation law $T^{AB}_{;B}=0$ gives:

\begin{equation}
\frac{1}{2} (p_r+ {\rho}) \nu'+p'_r+\frac{d-2}{r}(p_r-p_\theta ) =0 . \label{conservacion1}
\end{equation}

Note that the analogue Einstein tensor has free divergence $(\mathcal{G}^{(n)})^{AB}_{;B}=0$ \cite{Dadhich:2008df}, and the Bianchi Identities are satisfied \cite{Camanho:2015hea}. 

It is worth stressing that the $(d-2)$ angular components of the equations of motion are similar $(\theta,\theta)=(\phi,\phi)=...$ and the conservation equation can be written as a combination of the $(t,t)$,$(r,r)$ and $(\theta,\theta)$ components \cite{indio2}. In this way, there are three  field equations $(t,t),(r,r), (\theta,\theta)=(\phi,\phi)=...$ and one conservation equation \ref{conservacion1}. But only 3 equations are independent. Thus, any one equation could be ignored and  the system will be satisfied if the other three are solved.

\section{Pure Lovelock equations of motion for multiples sources} \label{seccion2}

In the equations of motion \ref{eqmovimiento}, $T^A_B=\mbox{diag}(-\rho,p_r,p_\theta,p_\theta,...)$ is given by equation \ref{EM}, and the seed energy momentum tensor is given by $\bar{T}^A_B= \mbox{diag}(-\bar{\rho},\bar{p}_r,\bar{p}_\theta,\bar{p}_\theta,...)$. 

In the Gravitational Decoupling method, reference \cite{Ovalle1}, developed for Einstein Hilbert theory, is introduced an additional source $(\theta_1)^A_B$ coupled with the seed energy momentum by the constant  $\alpha$. It is worth to notice that the power of $\alpha$ coincides with the value $n=1$ corresponding to the Einstein Hilbert theory. Thus, the energy momentum \ref{EM} is:

\begin{equation} \label{EMOvalle}
    T^A_B=\bar{T}^A_B+ \alpha (\theta_1)^A_B
\end{equation}
where the source $(\theta_1)^A_B$ is arbitrary. Thus, it is easily see that:

\begin{align}
    \rho&=\bar{\rho}- \alpha (\theta_1)^0_0 \label{densidadefectivaOvalle} \\
    p_r &= \bar{p}_r + \alpha (\theta_1)^1_1 \label{pradialefectivaOvalle} \\
    p_\theta &= \bar{p}_{\theta} + \alpha (\theta_1)^2_2 \label{ptangencialefectivaOvalle}
\end{align}

In this work, inspired by the above mentioned method, for a generic value of $n$, it is proposed the following  energy momentum tensor:
\begin{align} \label{EMMilko}
    T^A_B=&\bar{T}^A_B+ \alpha (\theta_1)^A_B + \alpha^2 (\theta_2)^A_B+...\nonumber \\
              &+\alpha^{n-1} (\theta_{n-1})^A_B+\alpha^n (\theta_{n})^A_B
\end{align}
therefore the number of sources is determined by the value of $n$. It is worth to stress that the energy momentum \ref{EMOvalle} is a particular case of \ref{EMMilko} for $n=1$. As example, for the Pure Gauss Bonnet case with $n=2$ the energy momentum \ref{EMMilko} has the form $T^A_B=\bar{T}^A_B+ \alpha (\theta_1)^A_B + \alpha^2 (\theta_2)^A_B$. Now:
\begin{align}
    \rho=&\bar{\rho}- \alpha (\theta_1)^0_0- \alpha^2 (\theta_2)^0_0 -... \nonumber \\
        & - \alpha^{n-1} (\theta_{n-1})^0_0 - \alpha^n (\theta_n)^0_0            \label{densidadefectiva} 
\end{align}
\begin{align}
    p_r =& \bar{p}_r + \alpha (\theta_1)^1_1 + \alpha^2 (\theta_2)^1_1 +... \nonumber \\
         & + \alpha^{n-1} (\theta_{n-1})^1_1 + \alpha^n (\theta_n)^1_1 \label{pradialefectiva} 
\end{align}
\begin{align} 
    p_\theta =&  \bar{p}_{\theta} + \alpha (\theta_1)^2_2 + \alpha^2 (\theta_2)^2_2+... \nonumber \\
              &+ \alpha^{n-1} (\theta_{n-1})^2_2 + \alpha^n (\theta_n)^2_2 \label{ptangencialefectiva}
\end{align}
where $p_\theta=p_\phi=...$ . In equation \ref{ptangencialefectiva} $\bar{p}_{\theta}=\bar{p}_{\phi}=...$ and we impose that $(\theta_i)^2_2=(\theta_i)^3_3=...$

So, for $(\theta_i)^1_1\neq (\theta_i)^2_2 =(\theta_i)^3_3=...$ and $\bar{p}_r \neq \bar{p}_\theta=\bar{p}_\phi=...$ these sources induce an anisotropy:

\begin{align}
    \Pi =& \bar{p}_\theta-\bar{p}_r+ \alpha \Big ( (\theta_1)^2_2-(\theta_1)^1_1 \Big ) + \alpha^2 \Big ( (\theta_2)^2_2-(\theta_2)^1_1 \Big ) \nonumber \\
         &+...+\alpha^{n-1} \Big ( (\theta_{n-1})^2_2-(\theta_{n-1})^1_1 \Big ) \nonumber \\
         &+ \alpha^{n} \Big ( (\theta_n)^2_2-(\theta_n)^1_1 \Big ). 
\end{align}
For the isotropic case where $\bar{p}_\theta=\bar{p}_r$, the addition of our source $\theta_{A B}$ is a simple way to generate an anisotropy.

So, the $(t,t)$ and $(r,r)$ components of the equations of motion are given by \cite{indio1,indio2}:

\begin{align} \label{tt}
&\frac{2}{d-2} r^{d-2}  \Big ( \bar{\rho}- \alpha (\theta_1)^0_0- \alpha^2 (\theta_2)^0_0 -... - \alpha^{n-1} (\theta_{n-1})^0_0  \nonumber \\
&- \alpha^n (\theta_n)^0_0 \Big) = \frac{d}{dr} \Big(r^{d-2n-1} \left( 1- e^{-\lambda}\right) ^n    \Big) ,
\end{align}
and
\begin{align} \label{rr}
&\frac{2}{d-2} r^{2n}\Big (\bar{p}_r + \alpha (\theta_1)^1_1 + \alpha^2 (\theta_2)^1_1 +...+ \alpha^{n-1} (\theta_{n-1})^1_1  \nonumber \\
&+ \alpha^n (\theta_n)^1_1 \Big) =nr\nu' e^{-\lambda}\left( 1- e^{-\lambda}\right)^{n-1} \nonumber \\
&-(d-2n-1)\left( 1- e^{-\lambda}\right)^n
\end{align}

We solve the $(t,t)$ and $(r,r)$ components of the Pure Lovelock equations together with the conservation equation. Using  the Bianchi identities, we ignore the remaining $(\theta,\theta)=(\phi,\phi)=...$ components (the suspense points indicate that all the tangential components of the Pure Lovelock equations are similar).

By inserting equations \ref{densidadefectiva},\ref{pradialefectiva} and \ref{ptangencialefectiva} into equation \ref{conservacion1}:

\begin{eqnarray} \label{conservacion2}
&&\frac{1}{2} (\bar{p}_r+\bar{\rho}) \nu'+\bar{p}_r'+ \frac{d-2}{r}(\bar{p}_r-\bar{p}_\theta ) \nonumber \\
&&+ \alpha \bigg ( \frac{1}{2} \Big ( (\theta_1)^1_1- (\theta_1)^0_0 \Big ) \nu'+\big ((\theta_1)^1_1 \big )'+\frac{d-2}{r}\Big ((\theta_1)^1_1-(\theta_1)^2_2 \Big ) \bigg ) \nonumber \\
&&+\alpha^2 \bigg ( \frac{1}{2} \Big ( (\theta_2)^1_1- (\theta_2)^0_0 \Big ) \nu'+\big ((\theta_2)^1_1 \big )'+\frac{d-2}{r} \Big ((\theta_2)^1_1-(\theta_2)^2_2 \Big ) \bigg ) \nonumber \\
&&+...+\alpha^{n-1} \bigg ( \frac{1}{2} \Big ( (\theta_{n-1})^1_1- (\theta_{n-1})^0_0 \Big ) \nu'+\big ((\theta_{n-1})^1_1 \big )' \nonumber \\
&&+\frac{d-2}{r} \Big ((\theta_{n-1})^1_1-(\theta_{n-1})^2_2 \Big ) \bigg ) \nonumber \nonumber \\
&& + \alpha^n \bigg ( \frac{1}{2} \Big ( (\theta_n)^1_1- (\theta_n)^0_0 \Big ) \nu'+\big ((\theta_n)^1_1 \big )'+\frac{d-2}{r}\Big ((\theta_n)^1_1-(\theta_n)^2_2 \Big ) \bigg ) \nonumber \\
&&=0 .
\end{eqnarray}
Thus, the system to solve corresponds to equations \ref{tt},\ref{rr} and \ref{conservacion2}. At this stage we would deal with five unknown functions, namely, the metric components $\lambda(r)$ and $\nu(r)$, and the effective thermodynamics functions $\rho$, $p_r$ and $p_\theta=p_\phi=...$. However, below, we implement the gravitational decoupling method where this scenario is modified.

The Pure Lovelock equations of motion for the seed energy momentum tensor are recovered for the limit $\alpha \to 0$ in the system \ref{tt}, \ref{rr} and \ref{conservacion2}. So, it is fulfilled that 
\begin{equation} \label{conservacionImpuesta1}
\nabla_A \bar{T}^A_B=0,
\end{equation}
and next, the first line of equation \ref{conservacion2} is conserved. For $n=1$ both components of energy momentum tensor are directly conserved, {\i.e} $\nabla_A(\theta_1)^A_B=0$, however, for $n>1$ one can notice that:
\begin{align}
&\alpha \nabla_A(\theta_1)^A_B + \alpha^2 \nabla_A(\theta_2)^A_B+...+\alpha^{n-1} \nabla_A(\theta_{n-1})^A_B \nonumber \\
&+\alpha^n \nabla_A(\theta_{n})^A_B=0,
\end{align}
where the covariant derivative is computed by using the line element \ref{metrica1}.  In this work we impose in arbitrarily way that:
\begin{equation} \label{conservacionImpuesta2}
 \alpha^i \nabla_A(\theta_i)^A_B=0.
  \end{equation}
 Thus, we will solve the system \ref{tt}, \ref{rr}, \ref{conservacionImpuesta1} and \ref{conservacionImpuesta2}. Under this assumption each source is separately conserved, and thus, there is no exchange of energy momentum between them. 
Therefore, our energy momentum tensor \ref{EMMilko} is a  way of decoupling the system inspired by the approach of reference \cite{Ovalle1}.

\section{ Gravitational decoupling by MGD in Pure Lovelock gravity} \label{metodo}
We start with a solution to equations \ref{tt}, \ref{rr}, \ref{conservacionImpuesta1} and \ref{conservacionImpuesta2} with $\alpha=0$, namely {\it seed solution} $\{\eta,\mu,\bar{\rho},\bar{p}_r,\bar{p}_t \}$, where $\eta$ and $\mu$ are the corresponding metric functions:

\begin{equation}
    ds^2=-e^{\eta(r)}dt^2+\mu(r)^{-1} dr^2+r^2 d\Omega^2_{d-2}. \label{metrica2}
\end{equation}

Turning on the parameter $\alpha$, the effects of the sources $(\theta_i)_{A B}$ appear on the seed solution $\{\eta,\mu,\bar{\rho},\bar{p}_r,\bar{p}_t \}$. These
effects can be encoded in the geometric deformation undergone by the seed fluid geometry $\{\eta,\mu \}$ in equation \ref{metrica2} as follows:
\begin{equation}
    \eta(r) \to \nu(r)=\eta(r) \label{deformaciontemporal}
\end{equation}

\begin{equation}
    \mu (r) \to e^{-\lambda} = \mu (r) - \alpha g(r). \label{deformacionradial}
\end{equation}

It means that only the radial component of the line element \ref{metrica2} is deformed, where $g(r)$ is the corresponding deformation of the radial part. This is known as {\it Minimal Geometric Deformation} \cite{Ovalle1} .Thus, replacing equations \ref{deformaciontemporal} and \ref{deformacionradial} into of equations \ref{tt} and \ref{rr}:

\begin{eqnarray} \label{tt1}
&&\frac{2}{d-2} r^{d-2}  \Big ( \bar{\rho}- \alpha (\theta_1)^0_0- \alpha^2 (\theta_2)^0_0 -... - \alpha^{n-1} (\theta_{n-1})^0_0  \nonumber \\
&&- \alpha^n (\theta_n)^0_0 \Big)=\frac{d}{dr} \Big (r^{d-2n-1} \big ( \left ( 1- \mu \right ) + \alpha g\big) ^n    \Big) ,
\end{eqnarray}
and
\begin{eqnarray} \label{rr1}
&&\frac{2}{d-2} r^{2n} \Big (\bar{p}_r + \alpha (\theta_1)^1_1 + \alpha^2 (\theta_2)^1_1 +...+ \alpha^{n-1} (\theta_{n-1})^1_1  \nonumber \\
&&+ \alpha^n (\theta_n)^1_1 \Big)=nr\nu' (\mu-\alpha g) \Big ( \left (1- \mu \right ) + \alpha g \Big)^{n-1} \nonumber \\
&&-(d-2n-1)\Big( \left (1- \mu \right ) + \alpha g \Big)^n .
\end{eqnarray}

Thus we must solve the system of equations \ref{tt1}, \ref{rr1}, \ref{conservacionImpuesta1} and \ref{conservacionImpuesta2}. We use the Binomial theorem :

\begin{align}
    (x+y)^N=&x^N +N x^{N-1}y + \left ( \begin{array}{c}  N\\ 2  \end{array} \right) x^{N-2}y^2+...+Nxy^{N-1} \nonumber \\
            &+ y^N,
\end{align}
thus:
\begin{eqnarray} \label{binomio1}
  \Big ( \left (1- \mu \right ) + \alpha g \Big)^n&=&\left (1- \mu \right )^n +n \left (1- \mu \right )^{n-1}g \alpha \nonumber \\
  &+& \left ( \begin{array}{c}  n\\ 2  \end{array} \right) \left (1- \mu \right )^{n-2}g^2\alpha^2+... \nonumber \\
  &+& n\left (1- \mu \right ) g^{n-1} \alpha^{n-1} + g^n \alpha^n,
\end{eqnarray}
and
\begin{eqnarray} \label{binomio2}
  \Big ( \left (1- \mu \right ) + \alpha g \Big)^{n-1}&=&\left (1- \mu \right )^{n-1} +(n-1) \left (1- \mu \right )^{n-2}g \alpha \nonumber \\
  &+& \left ( \begin{array}{c}  n-1\\ 2  \end{array} \right) \left (1- \mu \right )^{n-3}g^2\alpha^2+... \nonumber \\
  &+& (n-1)\left (1- \mu \right ) g^{n-2} \alpha^{n-2} + g^{n-1} \alpha^{n-1}, \nonumber \\
\end{eqnarray}

Thus, replacing equations \ref{binomio1} and \ref{binomio2} into of equations \ref{tt1} and \ref{rr1}, the system splits into the following sets of equations:

\begin{itemize}
\item The {\it standard Pure Lovelock equations} for a seed solution (with $\alpha=0$) :

\begin{equation} \label{ttcero}
\frac{2}{d-2} r^{d-2} \bar{\rho}  = \frac{d}{dr} \Big (r^{d-2n-1} \left( 1- \mu \right) ^n    \Big ) ,
\end{equation}
and
\begin{eqnarray} \label{rrcero}
&&\frac{2}{d-2} r^{2n}\left (\bar{p}_r \right)=nr\nu' \mu \left( 1- \mu \right)^{n-1} \nonumber \\
&&-(d-2n-1)\left( 1- \mu \right)^n
\end{eqnarray}

and the respective conservation equation:  

\begin{equation} \label{conservacioncero}
\frac{1}{2} (\bar{p}_r+\bar{\rho}) \nu'+\bar{p}_r'+ \frac{d-2}{r}(\bar{p}_r-\bar{p}_\theta )
=0
\end{equation}

\item The terms of order $\alpha$ give rise to the following { \it quasi-Pure Lovelock equations of order $\alpha^1$}, which include the source $\theta_{A B}$:
\begin{equation} \label{ttuno}
-\frac{2}{d-2} r^{d-2} (\theta_1)^0_0 = \frac{d}{dr} \Big (r^{d-2n-1}   n \left (1- \mu \right )^{n-1}g  \Big) , 
\end{equation}
\begin{align} \label{rruno}
    \frac{2}{d-2}r^{2n} (\theta_1)^1_1 =& n(1-\mu)^{n-1}g \bigg (r \nu' \Big ( (n-1) \mu  (1-\mu)^{-1}  \nonumber \\
                             &-1 \Big ) -(d-2n-1)    \bigg )
\end{align}

and the respective conservation equation: 
\begin{equation} \label{conservacionuno}
    \frac{1}{2} \Big ( (\theta_1)^1_1- (\theta_1)^0_0 \Big ) \nu'+\big ((\theta_1)^1_1 \big )'+\frac{d-2}{r}\Big ((\theta_1)^1_1-(\theta_1)^2_2 \Big )=0
\end{equation}

Thus, following the iteration, it is possible to obtain the quasi pure Lovelock equations of order $\alpha^2$, $\alpha^3$...$\alpha^{n-3}$, $\alpha^{n-2}$. 

\item The terms of order $\alpha^{n-1}$ give rise to the following { \it quasi-Pure Lovelock equations of order $\alpha^{n-1}$}:
\begin{equation} \label{ttn-1}
-\frac{2}{d-2} r^{d-2} (\theta_{n-1})^0_0  = \frac{d}{dr} \Big (r^{d-2n-1} n\left (1- \mu \right ) g^{n-1}    \Big ) ,
\end{equation}
and
\begin{align} \label{rrn-1}
    \frac{2}{d-2}r^{2n} (\theta_{n-1})^1_1 =& n g^{n-1} \Big ( r \nu' \big ( \mu- (n-1)(1-\mu)  \big ) \nonumber \\
                                 &- (d-2n-1)(1-\mu ) \Big )
\end{align}
and the respective conservation equation 
\begin{align} \label{conservacionn-1}
    &\frac{1}{2} \Big ( (\theta_{n-1})^1_1- (\theta_{n-1})^0_0 \Big ) \nu'+\big ((\theta_{n-1})^1_1 \big )' \nonumber \\
    &+\frac{d-2}{r} \Big ((\theta_{n-1})^1_1-(\theta_{n-1})^2_2 \Big ) =0 .
\end{align}

\item The terms of order $\alpha^{n}$ give rise to the following { \it quasi-Pure Lovelock equations of order $\alpha^{n}$}:
\begin{equation} \label{ttn}
- \frac{2}{d-2} r^{d-2} (\theta_n)^0_0  = \frac{d}{dr} \Big (r^{d-2n-1} g^n    \Big ) ,
\end{equation}
and
\begin{equation} \label{rrn}
    \frac{2}{d-2}r^{2n} (\theta_n)^1_1 = - g^n \Big (n r \nu' + (d-2n-1) \Big )
\end{equation}
and the respective conservation equation
\begin{equation} \label{conservacionn}
  \frac{1}{2} \Big ( (\theta_n)^1_1- (\theta_n)^0_0 \Big ) \nu'+\big ((\theta_n)^1_1 \big )'+\frac{d-2}{r}\Big ((\theta_n)^1_1-(\theta_n)^2_2 \Big ) =0.
\end{equation}

\end{itemize}

It is worth stressing that each quasi Pure Lovelock equation cannot be formally identified as the spherically symmetric Pure Lovelock equations for $n>1$, because the right sides of each quasi Pure Lovelock equation do not have the standard expressions for the Generalized Einstein tensor components $\mathcal{G}^{(n)}_{0 0}$ and $\mathcal{G}^{(n)}_{1 1}$. Furthermore, the Bianchi identities are not satisfied for each quasi Pure Lovelock equation. For $n=1$ the quasi Einstein equations can be transformed in the standard Einstein equations after a convenient redefinition of the energy momentum  tensor \cite{Ovalle1}, however, the method of the reference \cite{Ovalle1} has been widely used to find new solutions without using this mentioned redefinition in several works.

Despite the above mentioned, our imposed way for solving the system \ref{tt1}, \ref{rr1},  \ref{conservacionImpuesta1} and \ref{conservacionImpuesta2}, based in the decoupling of sources by means of the standard and quasi Pure Lovelock equations, ensures us to solve successfully the original system \ref{tt1}, \ref{rr1} and \ref{conservacion2}.
Furthermore, under our assumptions, each conservation equation \ref{conservacioncero}, \ref{conservacionuno},$...$  \ref{conservacionn-1}, \ref{conservacionn} is separately conserved, and thus, there is no exchange of energy momentum between the seed fluid and each sector $(\theta_i)_{A B}$. So, in our gravitational decoupling method there is only purely gravitational interaction. 

It is worth stressing that as a consequence of the application of the MGD:
\begin{itemize}
\item We start with the indefinite system \ref{tt},\ref{rr} and \ref{conservacion2}. After the application of MGD, we have a set of equations for the seed fluid $(\nu,\mu,\bar{\rho},\bar{p}_r,\bar{p}_\theta)$ given by the standard Pure Lovelock equations.

Next, we suppose that we have already found a seed fluid solution $(\nu,\mu)$ and the sources $(\bar{\rho},\bar{p}_r,\bar{p}_\theta)$, thus we have :

\item A much simpler system of four unknown functions $(g,(\theta_n)^0_0$
,$(\theta_n)^1_1$,$(\theta_n)^2_2)$ given by the quasi Pure Lovelock equations of order $\alpha^n$. 

\item Supposing that we have found the values of $g$ and $(\theta_n)_{AB}$, we have $n-1$ systems, given by the quasi Pure Lovelock equations of order $\alpha^i$, where each of them has three unknown functions $((\theta_i)^0_0,(\theta_i)^1_1,(\theta_i)^2_2)$
\end{itemize}

For the study of well behaved solutions that represent stellar distributions it is necessary to analyse the matching conditions \cite{Milko}. This is outside of the scope of this work and could be studied in elsewhere. 

\section{ A special case} \label{casoespecial}

We impose the condition $\nu=-\lambda$ in equation \ref{metrica1}. So, the $(t,t)$ component of Pure Lovelock equation keeps its form as in equation \ref{tt}. However, it is direct to check that the $(r,r)$ component, equation \ref{rr}, now is:

\begin{eqnarray} \label{rrCaso1}
&&\frac{2}{d-2} r^{d-2}\Big  (-\bar{p}_r - \alpha (\theta_1)^1_1 - \alpha^2 (\theta_2)^1_1 -...- \alpha^{n-1} (\theta_{n-1})^1_1 \nonumber \\ 
&&- \alpha^n (\theta_n)^1_1 \Big )  =\frac{d}{dr} \Big(r^{d-2n-1} \left( 1- e^{-\lambda}\right) ^n    \Big) ,
\end{eqnarray}
see reference \cite{indio3}. Now, the system to solve is : equation \ref{tt} that corresponds to $(t,t)$ component, equation \ref{rrCaso1} that corresponds to $(r,r)$ component, and equations \ref{conservacionImpuesta1} and \ref{conservacionImpuesta2}.

Furthermore, we impose the condition $\rho=-p_r$, as in references \cite{milko1,Dymnikova:2010zz}, where $\rho$ corresponds to equation \ref{densidadefectiva} and $p_r$ to equation \ref{pradialefectiva}. Additionally, we will impose arbitrarily that $\bar{\rho}=-\bar{p}_r$ and $(\theta_i)^0_0=(\theta_i)^1_1$, with $i=1,2...n$. Thus, the condition $\rho=-p_r$ is fulfilled.  

So, the $(t,t)$ and $(r,r)$ components are similar to equation \ref{tt}, whereas, now the conservation equation \ref{conservacion2} takes the following form:
\begin{align} \label{conservacion3}
&\bar{p}_r'+ \frac{d-2}{r}(\bar{p}_r-\bar{p}_t ) \nonumber \\
&+ \alpha \Big ( \big ((\theta_1)^1_1 \big )' +\frac{d-2}{r} \big ((\theta_1)^1_1-(\theta_1)^2_2 \big ) \Big ) \nonumber \\
&+\alpha^2 \Big ( \big ((\theta_2)^1_1 \big )' +\frac{d-2}{r} \big ((\theta_2)^1_1-(\theta_2)^2_2 \big ) \Big ) +...\nonumber \\
&+\alpha^{n-1} \Big ( \big ((\theta_{n-1})^1_1 \big )' +\frac{d-2}{r} \big ((\theta_{n-1})^1_1-(\theta_{n-1})^2_2 \big ) \Big )  \nonumber \\
&+\alpha^n \Big ( \big ((\theta_n)^1_1 \big )' +\frac{d-2}{r} \big ((\theta_n)^1_1-(\theta_n)^2_2 \big ) \Big )=0
\end{align}

In this way, the system to solve corresponds to the equations \ref{tt}, \ref{conservacionImpuesta1} with $\bar{\rho}=-\bar{p}_r$ and \ref{conservacionImpuesta2} with $(\theta_i)^0_0=(\theta_i)^1_1$. Our seed solution, which is solution of this system with $\alpha=0$, is:
\begin{equation}
    ds^2=-\mu(r)dt^2+\mu(r)^{-1} dr^2+r^2 d\Omega^2_{d-2}. \label{metrica3}
\end{equation}

Again, turning on $\alpha$, the effects of the source $\theta_{A B}$ appear on the seed solution. These
effects are encoded in the geometric deformation undergone by the seed fluid geometry in equation \ref{metrica3} as follows:
\begin{equation}
    \mu (r) \to e^{\nu} = \mu (r) - \alpha g(r). \label{deformaciontemporal1}
\end{equation}

\begin{equation}
    \mu (r) \to e^{-\lambda} = \mu (r) - \alpha g(r). \label{deformacionradial1}.
\end{equation}

Thus, taking into account the geometric deformation of equations \ref{deformaciontemporal1} and \ref{deformacionradial1}, the $(t,t)$ and $(r,r)$ components are similar to the equation \ref{tt1} and, the system to solve is given by equations \ref{tt1}, \ref{conservacionImpuesta1} with $\bar{\rho}=-\bar{p}_r$ and \ref{conservacionImpuesta2} with $(\theta_i)^0_0=(\theta_i)^1_1$.

Finally, using the binomial development \ref{binomio1}, the system splits into the following sets of equations: 

\begin{itemize}
    \item The standard Pure Lovelock equations for a seed solution (with $\alpha=0$), that correspond to the equations \ref{ttcero} and the conservation equation given by:
    \begin{equation} \label{conservacioncerocaso}
     \bar{p}_r'+ \frac{d-2}{r}(\bar{p}_r-\bar{p}_\theta )=0  
    \end{equation}
    \item The terms of order $\alpha$ give rise to the quasi-Pure Lovelock equations of order $\alpha^1$, that correspond to the equations \ref{ttuno} and the conservation equation given by:
    \begin{equation}
        \big ((\theta_1)^1_1 \big )' +\frac{d-2}{r} \big ((\theta_1)^1_1-(\theta_1)^2_2 \big )=0
    \end{equation}
Again, following the iteration, it is possible to obtain the quasi pure Lovelock equations of order $\alpha^2$, $\alpha^3$...$\alpha^{n-3}$, $\alpha^{n-2}$. 
\item The terms of order $\alpha^{n-1}$ give rise to the quasi-Pure Lovelock equations of order $\alpha^{n-1}$, that correspond to the equations \ref{ttn-1} and the conservation equation given by:
\begin{equation}
    \big ((\theta_{n-1})^1_1 \big )' +\frac{d-2}{r} \big ((\theta_{n-1})^1_1-(\theta_{n-1})^2_2 \big )=0
\end{equation}
\item The terms of order $\alpha^{n}$ give rise to the quasi-Pure Lovelock equations of order $\alpha^{n}$, that correspond to the equations \ref{ttn} and the conservation equation given by:
\begin{equation} \label{conservacionncaso}
    \big ((\theta_n)^1_1 \big )' +\frac{d-2}{r} \big ((\theta_n)^1_1-(\theta_n)^2_2 \big )=0 .
\end{equation}
\end{itemize}

\subsection{A very simple test: Applying the method to the AdS space time.} \label{RBH}

As a simple test, we apply our method to a seed Anti de Sitter space time and, we test if it is possible to obtain a solution of physical interest that represents {\it regular black holes}.

The Schwarzschild black hole has a singularity where the laws of physics cease to operate. From the classical point of view, Bardeen in reference \cite{Bardeen} proposed the first model of regular black holes, where the singularity is avoided, due to the formation of a dense core near the origin, whose effective cosmological constant $\Lambda_{eff}$ causes repulsive effects and, whose internal geometry has de Sitter form. This is achieved, by the change of the constant mass parameter by a mass function $M \to m(r)$ , such that near the origin $m(r) \approx \frac{\Lambda_{eff}}{6} r^3$, and $\displaystyle \lim_{r \to \infty} m(r)=M$. Thus, the factor $f=1- \frac{2m(r)}{r}$ near the origin behaves as de Sitter space time, and far from origin behaves as Scharzschild space time. After this, the formation of the Sitter core is associated to quantum fluctuations, where the energy density is of order of Planck units near the origin. This model is called {\it Planck star }, see references \cite{Planckstar1,Planckstar2}. Other more recent studies about regular black holes have been developed in references \cite{RBH1,RBH2,RBH3,RBH4,milko1}.

Due that regular black holes with de Sitter ground state have a cosmological horizon, we apply the method to a seed solution with Anti de Sitter structure. This is because, the presence of the cosmological horizon prevents a correct definition of the mass. Conversely, Anti de Sitter space time has a well defined asymptotically region, and thus, it is possible to define the mass. 

As was said above, Pure Lovelock is a theory that, for $\Lambda>0$, has a unique de Sitter ground state for $n$ odd, and has a double Anti de Sitter or de Sitter ground state for $n$ even. Conversely, for $\Lambda<0$, has a unique Anti de Sitter ground state for $n$ odd and does not have physical solution for $n$ even.

So, to find a seed solution with Anti de Sitter structure, we study the case where the seed fluid has the following energy density:
\begin{equation} \label{densidadpar}
    \bar{\rho}=-\bar{p}_r=\Lambda_d=\frac{(d-1)(d-2)}{2l^{2n}} \mbox{,     for $n$ even},
\end{equation}
and
\begin{equation} \label{densidadimpar}
    \bar{\rho}=-\bar{p}_r=\Lambda_d=-\frac{(d-1)(d-2)}{2l^{2n}} \mbox{,     for $n$ odd},
\end{equation}
thus, the energy density represents a $d$ dimensional cosmological constant, and $l^n$ is the Anti de Sitter radius. 

Then, we will apply the Gravitational Decoupling method as follows:

\begin{itemize}
    \item We solve the standard Pure Lovelock equations for our seed fluid:
    
First, we test the case with $n$ even. Inserting the seed energy density \ref{densidadpar} into equation \ref{ttcero}, we get to 
\begin{equation}
    \frac{r^{2n}}{l^{2n}}=(1-\mu)^n
\end{equation}
 where $d-2n-1>0$ and,whose solution is $\pm \frac{r^2}{l^2}=(1-\mu)$. Taking the minus sign:
\begin{equation} \label{musemilla}
    \mu=1+\frac{r^2}{l^2}
\end{equation}

For the case with $n$ odd, with $d-2n-1>0$, inserting the seed energy density \ref{densidadimpar} into equation \ref{ttcero}, it is direct  that the unique solution is equation \ref{musemilla}.

Thus, both for $n$ odd or $n$ even, our seed solutions represent an Anti de Sitter space time. The tangential seed pressure is determined by equation \ref{conservacioncerocaso}.
\item We solve the quasi-Pure Lovelock of order $\alpha^n$:

We choose a source $-(\theta_n)^0_0=F(r)$ as in reference \cite{milko1}: $F(r)$ has a maximum value at $r=0$, such that the $m(r)$ function near the origin behaves as :
\begin{equation} \label{condicionmasa}
    m(r)|_{r \approx 0} \approx \frac{r^{d-1}}{k^{2n}},
\end{equation}
where $k$ is a constant, and the $m(r)$ function is computed as:
\begin{equation}
   m(r)= \frac{2}{d-2} \int_0^r F(r) r^{d-2}dr,
\end{equation}

By solving equation \ref{ttn}:
\begin{equation}
    g(r)= \left ( \frac{m(r)}{r^{d-2n-1}} \right)^{1/n}
\end{equation}

So, our line element is given by
\begin{equation} \label{metricacaso}
    ds^2=-f(r) dt^2+\frac{dr^2}{f(r)}+r^2 d\Omega^2_{d-2}, 
\end{equation}
where:
\begin{equation} \label{solucion}
    f(r)= \mu - \alpha g = 1+\frac{r^2}{l^2}-\alpha \Big ( \frac{m(r)}{r^{d-2n-1}}   \Big )^{1/n},
\end{equation}
thus, near the origin the function behaves as:
\begin{equation}
    f(r)|_{r \approx 0}= 1 - \Big ( \frac{\alpha}{k^2}- \frac{1}{l^2}    \Big )r^2,
\end{equation}
thus, for 
\begin{equation} \label{condicion}
\alpha > \frac{k^2}{l^2},
\end{equation}
the solution behaves as de Sitter near the origin, and represents to a {\it Pure Lovelock regular black hole by gravitational decoupling}.

One example of $F(r)$ function is found in reference \cite{milko1}, wich is a $d$  dimensional generalization of Hayward density and, in our case is:
\begin{align}
    &F(r) = \frac{(d-1)(d-2)}{2} \frac{Q^{d-2}M^2}{(Q^{d-2}+r^{d-1})^2} \nonumber \\
    &\mbox{ which yields } m(r)=\frac{Mr^{d-1}}{Q^2M+r^{d-1}},
\end{align}
where, one can notice that:
\begin{align}
    &F(r)_{max}=F(0)= \frac{(d-1)(d-2)}{2 Q^{d-2}} \nonumber \\
    &\mbox{    and } m(r)|_{r \approx 0} \approx \frac{1}{Q^{d-2}}r^{d-1},
\end{align}
on the other hand:
\begin{equation}
\displaystyle \lim_{r \to \infty} F(r) = 0 \mbox{    and } \lim_{r \to \infty} m(r)= M,
\end{equation}
where $Q$ is defined in this reference as a regulator and $M$ is the total mass. There is a deep analysis of unities in reference \cite{milko1}.  So, $k^{2n}=Q^{d-2}$ and if the condition \ref{condicion} is satisfied, then the function $F(r)$ is suitable for represent a Pure Lovelock Regular black hole. The fact that $\displaystyle \lim_{r \to \infty} m(r)= M$ allows to define correctly the mass in a space time with AdS ground state.

The source $(\theta_n)^2_2$ is determined with equation \ref{conservacionncaso}

\item Known $\bar{\rho}, \bar{p}_r, \bar{p}_t, (\theta_n)^0_0, (\theta_n)^1_1 , (\theta_n)^2_2$, $\mu$, $\nu$ and $g$, the remaining sources $(\theta_i)^0_0, (\theta_i)^1_1 , (\theta_i)^2_2   $ are determined by the quasi-Pure Lovelock equations of order $\alpha^i$.
\end{itemize}

\section{Conclusion and discussion}
We have shown an approach that represents a simple method for decoupling gravitational sources in Pure Lovelock gravity. Thus, it is possible to decouple gravitational sources under the effects of higher curvature correction terms for space times with a number of dimensions greater than four.

Applying our method, the final solution obtained is the result of the decoupling of the Pure Lovelock equations in a seed sector described by the seed energy momentum tensor $\bar{T}^A_B$ and the quasi Pure Lovelock equations of order $\alpha^i$ described by the sources $(\theta_i)^A_B$. Thus, the equations of motion are solved for each sector separately and, by the superposition of these solutions, the complete solution is obtained.

The seed and the extra sources are separately conserved under the assumptions imposed in this work. The quasi Pure Lovelock equations are of order $\alpha^1,\alpha^2,...\alpha^{n-1},\alpha^n$. Therefore, the combination of these $n+1$ sectors only has gravitational interaction and does not have exchange of energy momentum. The order $\alpha^n$ corresponds to the order of the Generalized Einstein Tensor $\mathcal{G}^{(n)}_{AB}$. Thus, for $n=1$, where $\mathcal{G}^{(1)}_{AB}$ represents to the Einstein Tensor in Einstein Hilbert theory, the quasi Pure Lovelock are of order $\alpha^1$, and then, the quasi Einstein equations described in reference \cite{Ovalle1} are a particular case of our method. As indicates reference \cite{Ovalle1}, the quasi Einstein equations are useful to study the interaction between ordinary matter and the conjectured dark matter, so, our quasi Pure Lovelock equations perhaps could serve to investigate this problem in presence of higher curvature terms and in space times with a number of dimensions greater than four.

Furthermore, we have presented our method for the case where the line element has the form of equation \ref{metrica1} in section \ref{metodo}, and for the case where $\nu=-\lambda$ and $\rho=-p_r$ in section \ref{casoespecial}. As a simple test, we have applied our method to an Anti de Sitter seed solution. 
Choosing a source $F(r)$ that fulfills the condition \ref{condicionmasa}, we have found the {\it Pure Lovelock Regular black hole solution by gravitational decoupling} \ref{solucion}. This solution differs in its structure with the Pure Lovelock regular black hole found in reference \cite{milko1}, whose structure is:
\begin{equation} \label{solucionmilko}
    f(r)=1- \left (\frac{m(r)}{r^{d-2n-1}} \pm \frac{r^{2n}}{l^{2n}}   \right )^{1/n}.
\end{equation}
wich takes the sign $+(-)$ for $\Lambda>(<)0$.
Thus, equation \ref{solucion} is a new solution in Pure Lovelock gravity, and thus a new Pure Lovelock regular black hole. In our solution, the ground state is obtained with the seed fluid, and when the extra sources $(\theta_i)_{A B}$ are turn off. 
Although both solutions \ref{solucion} and \ref{solucionmilko} differ in its structure, both share the same structure of ground state : for $\Lambda >0$ both solutions have a double Anti de Sitter or de Sitter ground state for $n$ even (where we have chosen the AdS branch) , whereas for $\Lambda <0$ both solutions have a single Anti de Sitter ground state for $n$ odd. 

Although our solution \ref{solucion} is new by using Pure Lovelock gravity, one solution with similar mathematical structure was found by using the {\it $n$-fold degenerated ground state theory} in reference \cite{milko1}. Therefore both solutions share the horizons structure and thermodynamics features described in reference \cite{milko1}. However, both solutions have different ground state structure. The ground state in the $n$-fold degenerated ground state theory is obtained by choosing a suited election of the coupled constants $\gamma$ such that the solution has one unique Anti de Sitter ground state (or $n$ fold degenerated Anti de Sitter ground state) with $\Lambda<0$, whereas in our case the ground state structure is different and was described in the last paragraph.

So, we have showed that our Gravitational Decoupling method is a direct way to obtain regular black holes. In regarding this, the Minimal Geometric Deformation of the Anti de Sitter space time give rise to regular black holes, under the assumptions used in this work.

A simple recipe to apply our method could be:

\begin{itemize}
    \item Pick up a seed Pure Lovelock solution $\{\mu,\nu,\bar{\rho},\bar{p}_r, \bar{p}_t \}$ and solve the standard Pure Lovelock equations.
    \item Solve the quasi Pure Lovelock equations of order $\alpha^n$. In regarding this, we impose a form of the source $(\theta_n)^0_0$ and the function $g(r)$ is directly determined by the $(t,t)$ component. Furthermore, in direct way, the sources $(\theta_n)^1_1$ and $(\theta_n)^2_2$ are obtained by solving the $(r,r)$ and $(\theta,\theta)$ components, respectively.
    \item Well known $\bar{\rho}, \bar{p}_r, \bar{p}_t, (\theta_n)^0_0,(\theta_n)^1_1,(\theta_n)^2_2$, $\mu$, $\nu$ and $g$, the remaining functions $(\theta_i)^0_0,(\theta_i)^1_1,(\theta_i)^2_2$ are directly determined by the quasi-Pure Lovelock equations of order $\alpha^i$.
\end{itemize}

Thus, our method is an easy algorithm to search new analytical solutions of physical interest in Pure Lovelock gravity. In Einstein Hilbert theory, the Gravitational Decoupling method of reference \cite{Ovalle1} have been used to find new $4D$ black hole solutions in references \cite{Ovalle3,Contreras1,Contreras2,Contreras3,Contreras4} and new $4D$ well behaved solutions that represent stellar distributions in references \cite{Ovalle2,Camilo, Tello1, Graterol,Tello2,Tello3,Luciano,Milko, Luciano1}. So, inspired by this method, we have presented a useful tool that could serve to find new black hole solutions or stellar distributions in space times with a number of dimensions greater than four, and in presence of higher curvature correction terms. This applications could be studied in elsewhere.

\bibliography{mybib}

\end{document}